\documentclass{aa}
\usepackage{amssymb}
\usepackage{graphics}
\usepackage{times,float}

\begin{document}

\title{On the constraining observations of the dark \object{GRB 001109} and the properties of a $z$ $=$ 0.398 radio selected starburst galaxy contained in its error box\thanks{Based on observations made with telescopes at the Centro Astron\'omico Hispano  Alem\'an (1.23  m + 3.50  m), at the Observatorio del Roque  de los Muchachos (NOT + WHT), at the United States Naval Observatory (1.00 m)  and at the Russian Academy of  Sciences's Special Astrophysical Observatory (6.05 m). The NOT is operated on the island of San Miguel de la Palma jointly by Denmark, Finland, Iceland, Norway and Sweden, in Spain's Observatorio del Roque de los Muchahos of the Instituto de Astrof\'{\i}sica de Canarias. The Centro Astron\'omico Hispano  Alem\'an is operated in Calar Alto by the Max-Planck Institut f\"ur Astronomie of Heidelberg, jointly with Spain's Comisi\'on Nacional de Astronom\'{\i}a.}}
\titlerunning{The dark burst \object{GRB 001109}}

\author{J.M.   Castro Cer\'on       \inst{1,2}
   \and J.     Gorosabel            \inst{2,3,4}
   \and A.J.   Castro-Tirado        \inst{3}
   \and V.V.   Sokolov              \inst{5}
   \and V.L.   Afanasiev            \inst{5}
   \and T.A.   Fatkhullin           \inst{5}
   \and   S.N.~Dodonov              \inst{5}
   \and V.N.   Komarova             \inst{5}
   \and A.M.   Cherepashchuk        \inst{6}
   \and K.A.   Postnov              \inst{6}
   \and U.     Lisenfeld            \inst{3}
   \and J.     Greiner              \inst{7}
   \and S.     Klose                \inst{8}
   \and     J.~Hjorth               \inst{9}
   \and J.P.U. Fynbo                \inst{9,10}
   \and H.     Pedersen             \inst{9}
   \and E.     Rol                  \inst{11}
   \and J.     Fliri                \inst{12}
   \and M.     Feldt                \inst{13}
   \and G.     Feulner              \inst{12}
   \and M.I.   Andersen             \inst{14}
   \and   B.L.~Jensen               \inst{9}
   \and M.D.   P\'erez Ram\'{\i}rez \inst{15}
   \and F.J.   Vrba                 \inst{16}
   \and A.A.   Henden               \inst{16,17}
   \and G.     Israelian            \inst{18}
   \and N.R.   Tanvir               \inst{19}
   }

\offprints{J.M. Castro Cer\'on, \\
    \email{josemari@alumni.nd.edu}
    }

\institute{Real Instituto y Observatorio de la Armada, Secci\'on de Astronom\'{\i}a, 11.110 San Fernando-Naval (C\'adiz) Spain; {\tt josemari@alumni.nd.edu}.
      \and Space Telescope Science Institute, 3.700 San Mart\'{\i}n Dr., Baltimore MD 21.218--2.463 USA; {\tt gorosabel@stsci.edu}.
      \and Instituto de Astrof\'{\i}sica de Andaluc\'{\i}a (IAA-CSIC), Camino Bajo de Hu\'etor, 24, E--18.008 Granada Spain; {\tt jgu@iaa.es, ajct@iaa.es, ute@iaa.es}.
      \and Danish Space Research Institute, Juliane Maries Vej 30, DK--2\,100 K\o benhavn \O\ Denmark; {\tt jgu@dsri.dk}.
      \and Special Astrophysical Observatory of the Russian Academy of Sciences, Nizhnij Arkhyz, Karachai-Cherkessia, Russia 357\,147; {\tt sokolov@sao.ru, vafan@sao.ru, timur@sao.ru, dodo@sao.ru, vkom@sao.ru}.
      \and Sternberg Astronomical Institute, Mikhail Vasil'evich Lomonosov Moscow State University, Moscow, Leninskie Gory, Russia 119\,899; {\tt cher@sai.msu.su, moulin@sai.msu.su}.
      \and Max Planck Institut f\"ur extraterrestrische Physik, Giessenbachstra\ss e, Postfach 1\,312, D--85\,741 Garching Germany; {\tt jcg@mpe.mpg.de}.
      \and Th\"uringer Landessternwarte Tautenburg, D--07\,778 Tautenburg Germany; {\tt klose@tls-tautenburg.de}.
      \and Astronomical Observatory, University of Copenhagen, Juliane Maries Vej 30, DK--2\,100 K\o benhavn \O\ Denmark; {\tt jgu@astro.ku.dk, jens@astro.ku.dk, holger@astro.ku.dk, brian\_j@astro.ku.dk}.
      \and Institute of Physics and Astronomy, University of \AA rhus, Ny Munkegade, DK--8\,000 \AA rhus C Denmark; {\tt jfynbo@phys.au.dk}.
      \and Astronomical Institute Anton Pannekoek, University of Amsterdam, Kruislaan 403, NL--1\,098 SJ Amsterdam The Netherlands; {\tt evert@science.uva.nl}.
      \and Universit\"ats-Sternwarte M\"unchen, Scheinerstra\ss  e 1, D--81\,679 M\"unchen Germany; {\tt fliri@usm.uni-muenchen.de, feulner@usm.uni-muenchen.de}.
      \and Max Planck Institut f\"ur Astronomie, K\"onigstuhl 17, D--69\,117 Heidelberg Germany; {\tt mfeldt@mpia-hd.mpg.de}.
      \and Astrophysikalisches Institut, An der Sternwarte 16, D--14\,482 Potsdam Germany; {\tt mandersen@aip.de}.
      \and Sci-SA Department, ESTEC-ESA, Keplerlaan 1, Postbus 299, NL--2\,200 AG Noordwijk The Netherlands; {\tt dperez@rssd.esa.int}.
      \and US Naval Observatory, Flagstaff Station, P.O. Box 1.149, Flagstaff AZ 86.002-1.149 USA; {\tt fjv@nofs.navy.mil, ahenden@nofs.navy.mil}.
      \and Universities Space Research Association.
      \and Instituto de Astrof\'{\i}sica de Canarias, c/. V\'{\i}a L\'actea, s/n, E--38.200 La Laguna (Tenerife) Spain; {\tt gil@iac.es}.
      \and Department of Physical Sciences, University of Hertfordshire, College Lane, Hatfield, Herts GB--AL10 9AB UK; {\tt nrt@star.herts.ac.uk}.
      }

\date{Received 23 December 2003 / Accepted 4 June 2004}

\abstract{We present optical and NIR (near infrared) follow up observations of the \object{GRB 001109} from 1 to 300 days after the burst. No transient emission was found at these wavelengths within this GRB's (Gamma Ray Burst) 50\arcsec\ radius BeppoSAX error box. Strong limits (3$\sigma$) are set with: $R$ $\gtrsim$ 21, 10.2 hr after the GRB; $I$ $\gtrsim$ 23, 11.4 hr after the GRB; $H$ $\gtrsim$ 20.7, 9.9 hr after the GRB; and $K_S$ $\gtrsim$ 20, 9.6 hours after the GRB. We discuss whether the radio source found in the GRB's error box (\cite{taylor00}) might be related to the afterglow. We also present a multiwavelength study of a reddened starburst galaxy, found coincident with the potential radio and the \mbox{$X$ ray} afterglow. We show that our strong $I$ band upper limit makes of the \object{GRB 001109} the darkest one localised by the BeppoSAX's NFI (Narrow Field Instrument), and it is one of the most constraining upper limits on GRB afterglows to date. Further to it, the implications of these observations in the context of dark GRBs are considered.
}

\maketitle

\section{Introduction}

For the period spanning 1997--2001, approximately only one third of all GRBs (Gamma Ray Bursts) with well determined coordinates have had successful searches for optical counterparts (\cite{greiner03}). Several mechanisms (\cite{lazzati02}; \cite{ramírez-ruíz02}) have been presented to explain the lack of optical counterparts despite the prompt/deep observations carried out for some of them (\cite{fynbo01}; \cite{piro02}). It is thought that extinction around the progenitor and in the host galaxy plays a role in the non detection of the optical counterpart associated with dark GRBs (\cite{groot98}; \cite{taylor98}).

The \object{GRB 001109} was detected on 09.391169 UT November 2000 ($t_0$ hereafter) by the BeppoSAX (\cite{boella97}) with a refined uncertainty of 2.5\arcmin\ (\cite{gandolfi00ab}). A BeppoSAX NFI (Narrow Field Instrument) observation at $t_0$ + 16.5 hr detected a previously unknown source inside the 2.5\arcmin\ radius WFC (Wide Field Camera) error box (\cite{amati00}). The source, designated 1SAX J1830.1+5517, had R.A. (J2000) $=$ 18$^h$30$^m$07.8$^s$, Dec. (J2000) $=$ +55\degr17\arcmin56\arcsec\ (error radius $=$ 50\arcsec) and a 2--10 keV flux of 7.1 $\pm$ 0.5 $\times$ 10$^{-13}$ erg cm$^{-2}$ s$^{-1}$. See \cite{amati0304} for a detailed discussion of the \mbox{$X$ ray} properties.

A radio source (dubbed \mbox{\object{VLA J1830+5518}}) was found within the NFI error box with R.A.(J2000) $=$ 18$^h$30$^m$06.51$^s$, Dec.(J2000) $=$ +55\degr18\arcmin35.7\arcsec\ (conservative errors of 0.1\arcsec\ in each coordinate) and a flux of 236 $\pm$ 31 $\mu$Jy at 8.46 GHz (\cite{taylor00}). It seemed to decrease in brightness over a time span of 2 days (\cite{rol00}), but further observations at the VLA for $\sim$ 390 days failed to reveal a consistent decay (\cite{berger-frail01}).

In this paper we report on the deep optical/NIR (near infrared) observations carried out for the \object{GRB 001109} and their implications in the study of dark GRBs. Further we report on millimetre observations.

\begin{table*}
      \begin{center}
            \caption{Journal of observations of the \object{GRB 001109} field}
      \begin{tabular}{@{}lcccc@{}}

Date UT                  & Telescope              & Filtre      & Exposure Time             &  Limiting Magnitude \\
                         &                        &             & (seconds)                 &  (3$\sigma$)        \\

\hline
09.7708--09.8590/11/2000 & 1.23CAHA (CCD)         & $R$         &   7 $\times$    500       & 20.9$^{*~~}$   \\
09.7847--09.8854/11/2000 & 1.23CAHA (CCD)         & $B$         &   3 $\times$    600       & 20.3$^{*~~}$   \\
09.7848--09.7961/11/2000 & 4.20WHT  (INGRID)      & $K_S$       &                 750       & 19.9$^{**}$    \\
09.7968--09.8081/11/2000 & 4.20WHT  (INGRID)      & $H$         &                 750       & 21.0$^{**}$    \\
09.8083--09.8128/11/2000 & 4.20WHT  (INGRID)      & $J$         &                 300       & 21.3$^{**}$    \\
09.8447--09.8845/11/2000 & 2.56NOT  (StanCam)     & $I$         &   4 $\times$    600       & 22.9$^{~~~}$   \\
10.0876--10.1084/11/2000 & 1.00USNO (CCD)         & $I$         &              1\,800       & 21.0$^{*~~}$   \\
10.7363--10.7883/11/2000 & 3.50CAHA (OMEGA Prime) & $H$         &  10 $\times$    300       & 20.5$^{*~~}$   \\
10.7618--10.8417/11/2000 & 1.23CAHA (CCD)         & $R$         &   9 $\times$    500       & 20.9$^{*~~}$   \\
11.8191--11.8281/11/2000 & 4.20WHT  (INGRID)      & $H$         &                 600       & 20.7$^{**}$    \\
11.8292--11.8383/11/2000 & 4.20WHT  (INGRID)      & $K_S$       &                 600       & 19.4$^{**}$    \\
11.8423--11.8514/11/2000 & 4.20WHT  (INGRID)      & $J$         &                 600       & 21.4$^{**}$    \\
13.0560--13.0768/11/2000 & 1.00USNO (CCD)         & $I$         &              1\,800       & 21.0$^{*~~}$   \\
22.1590--22.1938/11/2000 & 2.56NOT  (ALFOSC)      & $B$         &                 600       & 23.0$^{~~~}$   \\
22.8278--22.8444/11/2000 & 2.56NOT  (ALFOSC)      & $U$         &   2 $\times$    600       & 24.0$^{~~~}$   \\
23.8035--22.8194/11/2000 & 2.56NOT  (ALFOSC)      & $B$         &   2 $\times$    600       & 23.5$^{~~~}$   \\
26.7576--26.7618/11/2000 & 3.50CAHA (MOSCA)       & $R$         &                 120       & 22.0$^{~~~}$   \\
27.7514--27.7556/11/2000 & 3.50CAHA (MOSCA)       & $R$         &                 180       & 22.3$^{~~~}$   \\
22.1590--22.1938/05/2001 & 4.20WHT  (PF)          & $B$         &   3 $\times$    900       & 24.0$^{~~~}$   \\
22.1951--22.2079/05/2001 & 4.20WHT  (PF)          & $V$         &   3 $\times$    450       & 23.5$^{~~~}$   \\
29.1249--29.1795/05/2001 & 2.56NOT  (ALFOSC)      & $U$         &   3 $\times$ 1\,500       & 23.5$^{~~~}$   \\
30.1249--30.1723/05/2001 & 2.56NOT  (ALFOSC)      & $V$         &                 900 + 300 & 23.5$^{~~~}$   \\
31.0468--31.0548/05/2001 & 2.56NOT  (ALFOSC)      & $V$         &                 600       & 22.0$^{~~~}$   \\
18.0361--18.0924/06/2001 & 4.20WHT  (PF)          & $U$         &   5 $\times$    900       & 23.5$^{~~~}$   \\
30.0583--30.1361/06/2001 & 3.50CAHA (OMEGA Cass)  & $K^\prime $ & 120 $\times$     60       & 21.0$^{\dag~}$ \\
01.0354--01.1181/07/2001 & 3.50CAHA (OMEGA Cass)  & $K^\prime $ & 120 $\times$     60       & 21.0$^{\dag~}$ \\
24.8655--24.8828/07/2001 & 6.05SAO  (SCORPIO)     & $R$         &   3 $\times$    180       & 25.5$^{~~~}$   \\
14.0524--14.0734/08/2001 & 2.56NOT  (ALFOSC)      & $R$         &                 600 + 900 & 23.8$^{~~~}$   \\
14.9983--15.0223/08/2001 & 2.56NOT  (ALFOSC)      & $R$         &   2 $\times$    900       & 24.0$^{~~~}$   \\
16.0571--16.1169/08/2001 & 2.56NOT  (ALFOSC)      & $B$         &   4 $\times$ 1\,200       & 25.0$^{~~~}$   \\
16.9835--17.0570/08/2001 & 2.56NOT  (ALFOSC)      & $U$         &   5 $\times$ 1\,500       & 24.1$^{~~~}$   \\
17.0148--17.0720/08/2001 & 2.56NOT  (ALFOSC)      & $V$         &   5 $\times$    900       & 24.5$^{~~~}$   \\
17.0720--17.1148/08/2001 & 2.56NOT  (ALFOSC)      & $I$         &   6 $\times$    600       & 23.7$^{~~~}$   \\
05.9503--06.0220/08/2002 & 2.20CAHA (BUSCA)       & $y$         &   6 $\times$    900       & 22.5$^{~~~}$   \\
\hline

\multicolumn{5}{l}{$^*$~~~\cite{greiner00}.~~~~~~Their $BRI$ band limiting magnitudes have been shifted to our zero point.}           \\
\multicolumn{5}{l}{$^{**}$~\cite{vreeswijk00}.~~Their $JHK_S$ band limiting magnitudes have been shifted to our zero point.}          \\
\multicolumn{5}{l}{$\dagger$~~~The images from 30/6--01/07/2001 were coadded in just a single limiting magnitude, $K^\prime$ $=$ 21.0.} \\

\hline

      \end{tabular}
            \label{registro de observaciones}
      \end{center}
\end{table*}

\begin{table*}
      \begin{center}
            \caption{Photometric secondary standards in the \object{GRB 001109} field}
      \begin{tabular}{@{}cccccccc@{}}

  & RA(J2000)   &Dec (J2000)              & $U$  & $B$              & $V$              & $R$              & $I$              \\
  & h~~~m~~~s   &\degr~~~\arcmin~~~\arcsec&      &                  &                  &                  &                  \\

\hline
1 & 18 29 52.55 & +55 16 37.8 & 18.62 $\pm$ 0.03 & 18.53 $\pm$ 0.08 & 17.95 $\pm$ 0.02 & 17.58 $\pm$ 0.02 & 17.25 $\pm$ 0.02 \\
2 & 18 30 18.61 & +55 16 46.6 & 21.02 $\pm$ 0.17 & 19.57 $\pm$ 0.04 & 18.49 $\pm$ 0.02 & 17.79 $\pm$ 0.02 & 17.16 $\pm$ 0.02 \\
3 & 18 30 02.94 & +55 17 03.2 & 19.24 $\pm$ 0.06 & 18.48 $\pm$ 0.07 & 17.55 $\pm$ 0.02 & 16.90 $\pm$ 0.02 & 16.36 $\pm$ 0.02 \\
4 & 18 30 04.05 & +55 17 33.7 & 21.31 $\pm$ 0.17 & 19.99 $\pm$ 0.05 & 18.97 $\pm$ 0.02 & 18.16 $\pm$ 0.02 & 17.52 $\pm$ 0.02 \\
5 & 18 29 48.91 & +55 19 20.5 & 20.26 $\pm$ 0.12 & 19.27 $\pm$ 0.06 & 18.33 $\pm$ 0.02 & 17.73 $\pm$ 0.02 & 17.25 $\pm$ 0.02 \\
6 & 18 30 22.09 & +55 19 36.9 & 19.16 $\pm$ 0.06 & 19.45 $\pm$ 0.02 & 19.03 $\pm$ 0.07 & 18.77 $\pm$ 0.02 & 18.45 $\pm$ 0.02 \\
7 & 18 30 20.65 & +55 19 40.7 & 20.63 $\pm$ 0.15 & 20.49 $\pm$ 0.08 & 20.05 $\pm$ 0.02 & 19.68 $\pm$ 0.03 & 19.27 $\pm$ 0.04 \\
8 & 18 30 14.57 & +55 20 43.3 & 19.54 $\pm$ 0.08 & 18.62 $\pm$ 0.03 & 17.28 $\pm$ 0.02 & 16.33 $\pm$ 0.02 & 15.30 $\pm$ 0.02 \\
\hline

      \end{tabular}
            \label{estrellas secundarias}
      \end{center}
\end{table*}

\section{Observations}

\begin{figure}
      \resizebox{\hsize}{!}{\includegraphics{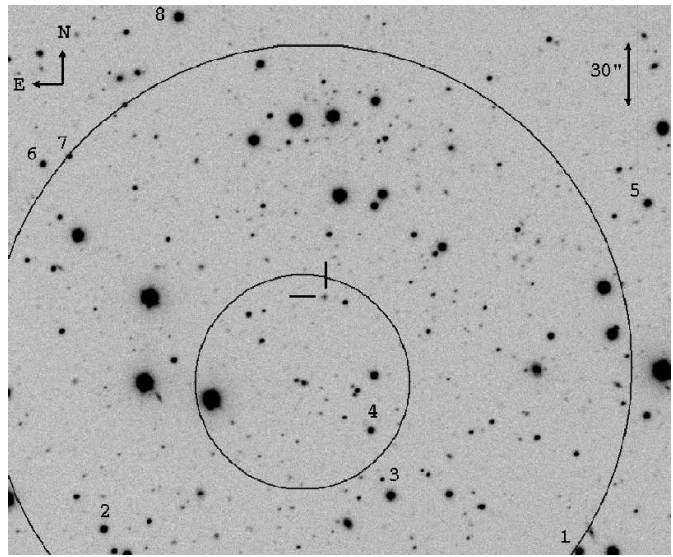}}
                 \caption{The contents of the BeppoSAX error box for the \object{GRB 001109} field. This $R$ band image was taken with the 2.56NOT (+ALFOSC) on 14.0524--14.0734 UT August 2001. The source in between ticks is the galaxy coincident with the \mbox{\object{VLA J1830+5518}} and consistent with the position of the \mbox{$X$ ray} afterglow. The numbered stars are the secondary standards indicated in Table \ref{estrellas secundarias}. The large circle represents the refined WFC error box (\cite{gandolfi00b}) and the small one the NFI error circle (\cite{amati00}). The field of view covered by the figure is 5.1\arcmin\ $\times$ 4.3\arcmin. North is upwards and East is leftwards.}
                 \label{zona de error}
\end{figure}

Table \ref{registro de observaciones} displays the observing log. Target of Opportunity observations started at $t_0$ + 9.1 hr (referred to the start time of the first frame reported by \cite{greiner00}, taken with the 1.23CAHA). We performed aperture photometry using SExtractor\footnote{http://terapix.iap.fr/soft/sextractor/} (\cite{bertin-arnouts96}) to study the contents of the BeppoSAX error box. The field was calibrated observing the Landolt field \object{SA113} (\cite{landolt92}) in the $UBVRI$ bands ($R$ and $I$ in the Cousins system), at airmasses similar to that of the GRB's field, in only one night. Table \ref{estrellas secundarias} shows the positions and magnitudes of several secondary standards in the GRB's field (see Fig. \ref{zona de error}). Spectroscopic observations were made at the 6.05SAO telescope (12 $\times$ 600 s exposures; see Fig. \ref{espectro}) with SCORPIO and a 300 lines/mm grating. The spectral resolution (FWHM) obtained was $\sim$ 20 \AA\ and the effective wavelength coverage was 3\,500--9\,500 \AA\ (\cite{afanasiev01}). Millimetre observations were carried out at the 30 m IRAM telescope (see Sect. \ref{resultados de las observaciones milimétricas}).

\section{Results and analysis}

\subsection{Content of the BeppoSAX NFI error box}

No optical afterglow was detected in the first 1.23CAHA (\cite{greiner00}; $R_{lim}$ $>$ 20.9 mag at 10.2 hr after the GRB) and 2.56NOT ($I_{lim}$ $>$ 22.9 mag at 11.4 hr after the GRB) frames. Strong limits come from the deep NIR observations. The $H$ and $K^\prime$ 3.50CAHA images (\cite{greiner00} reported the value of $H$) have been compared to the $H$ and $K_S$ 4.20WHT ones reported by \cite{vreeswijk00}. We derived the following upper limits\footnote{We have assumed $K^\prime$ $\simeq$ $K_S$} for any NIR transient emission within the NFI error box: $K^\prime >$ 19.9, $H >$ 20.7 and $J >$ 21.3, $\sim$ 10 hr after the GRB, all of them with a 3$\sigma$ confidence level.

\begin{figure}
      \resizebox{\hsize}{!}{\includegraphics{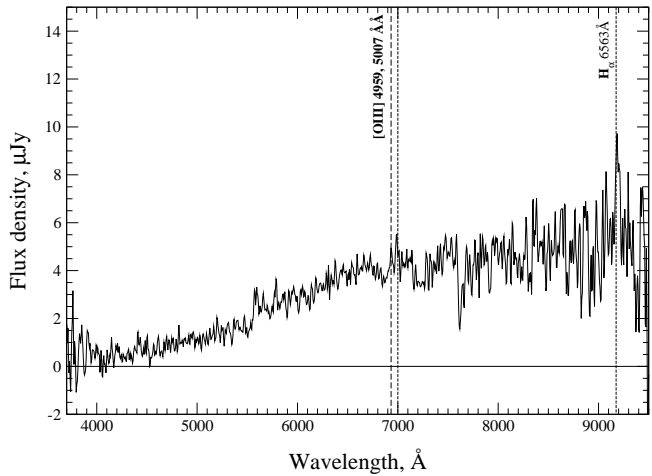}}
                 \caption{Optical spectrum of the galaxy coincident with the \mbox{\object{VLA J1830+5518}}, obtained with the 6.05SAO (+SCORPIO). It shows rest frame emission lines for H$\alpha$ (6\,563 \AA) and O[III] (4\,959 \AA, 5\,007 \AA).}
                 \label{espectro}
\end{figure}

\subsection{Afterglow's SED}
        \label{distribución de energía espectral de la contrapartida - sección}

\begin{figure}
      \resizebox{\hsize}{!}{\includegraphics{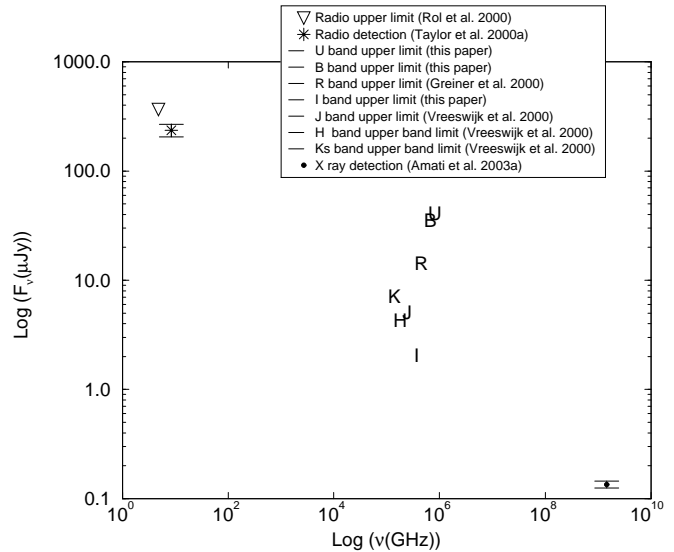}}
                 \caption{Selected detections (\mbox{$X$ ray}) and upper limits ($UBVRIJHK_S$) associated with the \object{GRB 001109}. The most constraining upper limit corresponds to the 2.56NOT $I$ band measurement.}
                 \label{distribución de energía espectral de la contrapartida - figura}
\end{figure}

Fig. \ref{distribución de energía espectral de la contrapartida - figura} displays selected detections and upper limits associated with the \object{GRB 001109} (to keep the figure legible we have only plotted the most constraining measurement for each of the optical and NIR bands). All plotted measurements have been shifted to a common epoch [$t$ $=$ $t_0 +$ 0.4 days; epoch of the radio detection (\cite{taylor00})] assuming a power law decay index $\alpha$ $=$ 1.27 (suggested by the \mbox{$X$ ray} observations reported in \cite{amati03}). As shown, the most constraining upper limit corresponds to the 2.56NOT $I$ band image taken on 9.8447--9.8845 UT November 2000.

This $I$ band measurement allows us to impose an upper limit on the afterglow's optical to \mbox{$X$ ray} spectral index: \mbox{$\beta_{optical-X ray}$} $<$ 0.33 $\pm$ 0.02 ($\beta$ is the power law index of the specific flux; $F_\nu$ $\sim$ $\nu^{-\beta}$). The corresponding optical to \mbox{$X$ ray} spectral index upper limits for the full BeppoSax's NFI dark burst sample can be worked out from the limits on the optical to \mbox{$X$ ray} flux ratio\footnote{The optical to \mbox{$X$ ray} flux ratio ($f_{oX}$) is defined as the $R$ band flux (or upper limit), in units of $\mu$Jy, divided by the 1.6--10 keV \mbox{$X$ ray} flux, in units of 10$^{-13}$ cgs (\cite{depasquale03}).} ($f_{oX}$ from Table 1 in \cite{depasquale03}):

$$f_{oX} = \frac{\nu_{optical}^{-\beta}}{\nu_X^{-\beta}}$$

The \object{GRB 001109} has the strongest limit in this sample where the upper limits range from 0.33 to 0.55.

\subsection{\mbox{\object{VLA J1830+5518}}}

\subsubsection{Astrometry}

\begin{figure}
      \resizebox{\hsize}{!}{\includegraphics{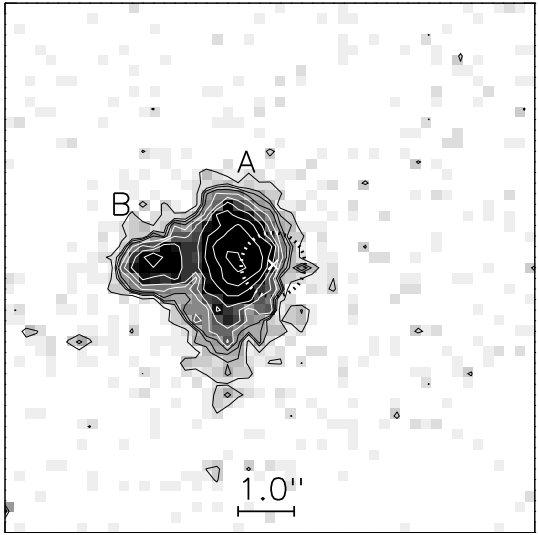}}
                 \caption{Contour plot displaying the galaxy coincident with the \mbox{\object{VLA J1830+5518}}. The figure shows the coadded I band image taken with the 2.56NOT (+ALFOSC) (17.0720--17.1148/08/2001) of the source coincident with the \mbox{\object{VLA J1830+5518}}. A seeing of 0.8\arcsec\ allowed us to separate the two components, the objects A and B. The centre of the circle marks the position of the radio source (\cite{taylor00}), R.A.(J2000) $=$ 18$^h$30$^m$06.51$^s$, Dec.(J2000) $=$ 55\degr18\arcmin35.7\arcsec. The radius of the circle is 0.57\arcsec. The contours show the detection confidence level above the background in a logarithmic scale. North is upwards and East is leftwards.}
                 \label{astrometría}
\end{figure}

We performed astrometry on two different data sets. For the first data set, 10 USNO A2.0 stars, not saturated on the 6.05SAO images, were used. The astrometrical uncertainty was found to be $\sim$ 0.5\arcsec, including both, statistical and systematic errors (\cite{sokolov01}). For the second data set, an independent astrometric solution, based on 50 USNO A2.0 stars, was obtained using the coadded $I$ band image taken at the 2.56NOT (see the penultimate entry in Table \ref{registro de observaciones}). It yielded a similar uncertainty (0.57\arcsec). Both astrometric solutions showed, independently, that the position of the radio source is consistent with the brighter component ($R$ $=$ 20.65 $\pm$ 0.06, the object A hereafter) of a complex system (see Fig. \ref{astrometría}). The second brightest component (the object B hereafter) is located 1.25\arcsec\ to the East of the object A. Strictly speaking the astrometry reveals the registration of the optical image, and has nothing to do with the location of the VLA radio source which was determined absolutely in the International Coordinate Reference Frame. These two objects were independently detected in the optical and in the NIR, so we conclude that they are real objects.

\subsubsection{Spectroscopy}

Spectral measurements (performed with the slit aligned in the East-West direction) detected Balmer breaks and emission lines for the sources A and B. First we divided the 2D spectra of the objects A and B using Gaussian analysis. Then we aproximated the resulting 2D spectrum across the dispersion direction by summing the two gaussians with the FWHM parametres (wavelength dependent). Finally we checked our extraction model by subtracting the model from the real data. Object A's redshift is $z$ $=$ 0.398 $\pm$ 0.002 based on the identification of the H$\alpha$ (6\,563 \AA) and O[III] (4\,959 \AA, 5\,007 \AA) emission lines (see Fig. \ref{espectro} and \cite{afanasiev01}). Object B's redshift is $z$ $=$ 0.3399 $\pm$ 0.0005 based on the identification of the H$\alpha$ (6\,563 \AA) and H$\beta$ (4\,861 \AA) emission lines (\cite{afanasiev01}). The redshift difference between sources A and B corresponds to a large relative expansion velocity of $\sim$ 13\,000 km s$^{-1}$. Given that velocity dispersions in galaxy clusters are, at most, $\sim$ 5\,000 km s$^{-1}$ (\cite{fadda96}; \cite{girardi93}), the alignment of both sources is likely the result of a chance projection. An HST high resolution deep image that might find signs of interaction would help to clarify this issue.

If the object A were the host of the \object{GRB 001109}, then the burst redshift would be $z$ $=$ 0.398 $\pm$ 0.002. To further constrain this suggestion we calculate the probability to find a radio source with the brightness of \mbox{\object{VLA J1830+5518}} in an error box with a radius of 50\arcsec.

Following \cite{fomalont02} the density of radio sources detected at 8.4 GHz above a flux density $S$ in microjanskys is given by:

$$N = (0.099 \pm 0.010) (\frac{S}{40})^{-1.11 \pm 0.13}~~~arcmin^{-2}$$

So we conclude that the chance probability of having a source brighter than $238 \pm 31 \mu Jy$ inside the NFI error box is $3 \pm 0.9$\% and thus, consider that the probability is not low enough to establish a physical relationship between the location of this radio source inside the GRB's error box and the occurrence of the \mbox{$\gamma$ ray} event.

\subsection{Object A's SED}

We have determined the flux distribution of the galaxy coincident (not necessarily related) with the \mbox{\object{VLA J1830+5518}} by means of our $UBVRI$ broad band photometric measurements together with the $JHK_S$ broad band measurements reported by \cite{vreeswijk00}. The photometry was based on SExtractor (\cite{bertin-arnouts96}), which allows to deblend entangled sources (this is specially relevant for cases like that of the objects A and B). The fluxes at the $UBVRIJHK_S$ passband wavelengths have been dereddened of Galactic extinction using a value of E ($B-V$) $=$ 0.04 (DIRBE/IRAS dust maps: \cite{schlegel98}). 

The $UBVRIJHK_S$ passband fluxes (measured in units of 2 $\times$ 10$^{-17}$ erg cm$^{-2}$ s$^{-1}$ \AA$^{-1}$; see Fig. \ref{distribución de energía espectral de la galaxia}) correspond to the following values: 0.118 $\pm$ 0.014, 0.234$ \pm$ 0.007, 0.475 $\pm$ 0.012, 0.684 $\pm$ 0.034, 0.573 $\pm$ 0.030, 0.526 $\pm$ 0.023, 0.393 $\pm$ 0.018 and 0.315 $\pm$ 0.014, respectively. We have modelled the SED using stellar population synthesis techniques (\cite{bolzonella00}) leaving the extinction and the redshift as free parametres. For the extinction law we have used the one given by \cite{calzetti00}, which is typical for starburst galaxies. The best fit is obtained with a dusty galaxy SED at $z$ $=$ 0.381, with A$_V$ $=$ 1.4 mag and an episode of star formation 0.25 Gyr ago (see Fig. \ref{distribución de energía espectral de la galaxia}). This episode of star formation dominates the optical continuum.

Although the radio emission from the object A is not related to the afterglow of the GRB, the object A could still be the host galaxy of the burst. To further constrain this suggestion we have calculated the number of galaxies, with magnitude $B$ $<$ 22.96 (reddened), to be found inside a circular area with a 50\arcsec\ radius. Using the Millennium Galaxy Catalogue (\cite{liske03}) we estimate a count of $\sim$ 3 galaxies (the $B$ passband flux has been dereddened of Galactic extinction using a value of E ($B-V$) $=$ 0.04; DIRBE/IRAS dust maps: \cite{schlegel98}).

The estimated extinction (A$_V$ $=$ 1.4) might explain the lack of optical emission and agree with the presence of the H$\alpha$ (6\,563 \AA) and O[III] (4\,959 \AA, 5\,007 \AA) emission lines (see Fig. \ref{espectro} and \cite{afanasiev01}) and the non negligible intrinsic N$_H$ $=$ 2.83 $\times$ 10$^{22}$ $\pm$ $^{4.7}_{2.83}$ cm$^{-2}$ (\cite{depasquale03}). This would also be consistent with the fact that the majority of the long duration GRB afterglows located so far have been linked to actively star forming galaxies.

\subsection{Results of the millimetre observations}
     \label{resultados de las observaciones milimétricas}

We observed the \mbox{\object{VLA J1830+5518}} with the 117 channel Max Planck Millimetre Bolometre array (MAMBO, \cite{kreysa98}) at the IRAM's 30 m radiotelescope on Pico Veleta, Spain, between 4 Mar 2003 and 12 Mar 2003. MAMBO has an effective centre frequency of $\sim$ 250 GHz (1.2 mm) and a beam size of 10.6\arcsec. The observations were done in standard on/off mode with 2 Hz chopping of the secondary mirror with a throw of 32\arcsec. The flux was calibrated by performing observations on Mars and Uranus, which yielded a conversion factor of 30\,000 counts/Jy with an estimated error of 15\%. We did not detect any emission from the \mbox{\object{VLA J1830+5518}} down to a rms noise level of 0.5 mJy.

\begin{figure}
      \resizebox{\hsize}{!}{\includegraphics{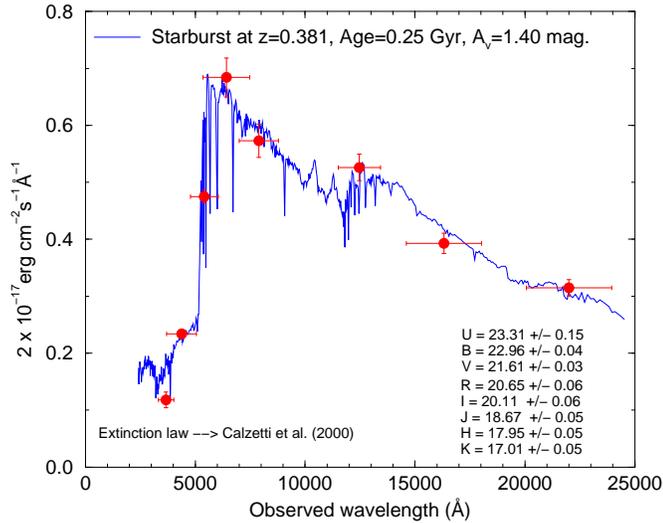}}
                 \caption{SED of the galaxy coincident with the \mbox{\object{VLA J1830+5518}}. The solid line is the SED of a dusty galaxy at $z$ $=$ 0.381, with A$_V$ $=$ 1.4 mag and an episode of star formation 0.25 Gyr ago. This episode of star formation dominates the optical continuum. The SED has been constructed with homogeneous data taken with the 2.56NOT (+ALFOSC) for the optical and with the 4.20WHT (+INGRID) for the NIR.}
                 \label{distribución de energía espectral de la galaxia}
\end{figure}

Further, we have estimated how ``unusual'' is a non detection with MAMBO in this case. This estimation is based in a correlation between the far IR and the radio. \cite{carilli-yun99} give the correlation of the far IR/radio bands as a function of the redshift. Adapting this correlation to our frequencies we expect, for $z$ $=$ 0.398, the fluxes at 8.4 GHz and 250 GHz to be aproximately equal. We have that, for 8.4 GHz, flux $=$ 0.2 mJy, and that, for 250 GHz, flux $<$ 1.5 mJy (3$\sigma$), so our results agree with the expected ones. With a higher redshift the flux at 250 GHz is expected to rise as a function of the flux in radio. Our upper limit at 250 GHz and the correlation from \cite{carilli-yun99} give us an estimate of $z$ $=$ 1 as the maximum redshift allowed for the VLA source (taking the flux at 250 GHz to be less than 1.5 mJy at 3$\sigma$). This consistency with the far IR/radio correlation implies that the radio emission probably originates from star formation and not from an AGN. Such conclusion can be accomodated by our $UBVRIJHK$ band SED of the object A (see Table \ref{magnitudes del objeto A} and Fig. \ref{distribución de energía espectral de la galaxia}). Additionally, the $z$ $=$ 1 upper limit for the \mbox{\object{VLA J1830+5518}} is consistent with the object A's redshift.

\mbox{\cite{barnard03} observed} the \mbox{\object{VLA J1830+5518}} with the $\sim$ 350 GHz photometry pixel on the Submillimetre Common User Bolometre Array (SCUBA, \cite{holland99}), at the James Clerk Maxwell Telescope on Mauna Kea, United States. Their measurement yielded a flux of 1.89 $\pm$ 1.4 mJy, consistent with our upper limit at 250 GHz since the flux at 350 GHz is larger by a factor of 3--4 (considering a rms noise level of 0.5 mJy).

\begin{table}
      \begin{center}
            \caption{The object A's magnitudes}
      \begin{tabular}{@{}cc@{}}

Band & Magnitude        \\

\hline
 $U$ & 23.31 $\pm$ 0.15 \\
 $B$ & 22.96 $\pm$ 0.04 \\
 $V$ & 21.61 $\pm$ 0.03 \\
 $R$ & 20.65 $\pm$ 0.06 \\
 $I$ & 20.11 $\pm$ 0.06 \\
 $J$ & 18.67 $\pm$ 0.05 \\
 $H$ & 17.95 $\pm$ 0.05 \\
 $K$ & 17.01 $\pm$ 0.05 \\
\hline

      \end{tabular}
            \label{magnitudes del objeto A}
      \end{center}
\end{table}

\section{Discussion and conclusions}

Our optical/NIR/millimetre observations are consistent with a connection between the \mbox{\object{VLA J1830+5518}} and the object A. On the other hand, a connection between the object A and the \object{GRB 001109} can not be established.

If we define a dark GRB as one with no counterpart brighter than $R$ $<$ 23 at 24 hr from the onset, then the \object{GRB 001109} is clearly a dark GRB, given the $I$ $>$ 22.9 limit imposed by the 2.56NOT 0.47 days after the \mbox{$\gamma$ ray} event. We have used our 2.56NOT $I$ $>$ 22.9 upper limit (9.8447--9.8845/11/2000 UT) to further constrain the luminosity of the \object{GRB 001109} within the context of the BeppoSAX's NFI dark burst sample (see upper limits for dark bursts in Table 1 of \cite{depasquale03}). To do so we have (following the methodology described in \cite{depasquale03}) calculated the $R$ band upper limit 11 hr after the GRB, from the 2.56NOT $I$ band constraint. First, we calculated the $R$ band flux associated with the $I$ band limit using the spectral index \mbox{$\beta_{optical-X ray}$} $=$ 0.33 $\pm$ 0.02 (see Sect. \ref{distribución de energía espectral de la contrapartida - sección}). Then, the $R$ band flux was rescaled from $t-t_0$ $=$ 11.3624 hr (9.8646 UT November 2000; mean 2.56NOT observing time) to $t-t_0$ $=$ 11 hr (assuming a power law decay index $\alpha$ $=$ 1.15, adopted by \cite{depasquale03}). Further, the $R$ band flux upper limit was corrected for Galactic extinction using a E ($B-V$) $=$ 0.04 value (\cite{schlegel98}). As a result we derived an unextincted $R$ band flux upper limit of 1.80 $\mu$Jy 11 hr after the \mbox{$\gamma$ ray} event. This new $R$ band flux upper limit is approximately 7 times deeper than the one reported previously (11.81 $\mu$Jy in \cite{depasquale03}). Moreover, our $I$ band image lowers the $f_{oX}$ from 0.59 to 0.09, making the \object{GRB 001109}, by far, the darkest BeppoSAX NFI GRB. Consequently, the 2.56NOT $I$ band measurement has impossed one of the most constraining upper limits on GRB afterglows to date.

The \object{GRB 001109} belongs to the subsample of darkest BeppoSAX NFI bursts ($\sim 25\%$ of the total BeppoSAX NFI dark GRB sample) which show $f_{oX}$ values incompatible (at a 2.6$\sigma$ level) with GRBs with detected optical transients (\cite{depasquale03}). For those objects the spectral index \mbox{$\beta_{optical-X ray}$} $\le$ 0.62, so the \object{GRB 001109} is clearly in this group, which is composed by the GRBs \object{981226}, \object{990704}, \object{990806} and \object{000210}.

It is important to highlight that the \object{GRB 001109} exhibited the brightest \mbox{$X$ ray} afteglow among the dark BeppoSAX NFI bursts (\cite{depasquale03}). On the other hand it showed the lowest N$_H$ reported for the dark BeppoSAX NFI GRBs (\cite{depasquale03}). In fact, the \object{GRB 001109} N$_H$ value is consistent with the ones measured for GRBs with detected optical transients. Thus, the bright \mbox{$X$ ray} afterglow of the \object{GRB 001109}, its low N$_H$ value (in comparison to the rest of the dark BeppoSAX NFI GRB sample) and the constraining optical limits imposed in the present work, might indicate that the \object{GRB 001109} showed a spectrum intrinsically different from GRBs with detected optical transients.

\section*{Acknowledgments}

For their assistance during the observations we are grateful to, L. Montoya and A. Aguirre at the CAHA, A.A. Kass at the NOT and, I. Skillen, C. Benn, T. Augusteijn and C. Zurita at the WHT. Some images were taken under the auspices of the Isaac Newton Group's Service Observation Programme. Some of the data presented here have been taken using ALFOSC, which is owned by the Instituto de Astrof\'{\i}sica de Andaluc\'{\i}a (IAA-CSIC) an operated at the Nordic Optical Telescope under agreement between the IAA-CSIC and the NBIfAFG of the Astronomical Observatory of Copenhagen. We thank E. P\'erez Jim\'enez and M.A. Cervi\~no Saavedra for fruitful discussions. JMCC and JG acknowledge a FPI doctoral fellowship from Spain's Ministerio de Ciencia y Tecnolog\'{\i}a and a Marie Curie Research Grant from the European Commission respectively. JMCC is grateful for the hospitality extended by the Danish Space Research Institute (DSRI), the Instituto de Astrof\'{\i}sica de Andaluc\'{\i}a (IAA-CSIC) and the Astronomical Observatory of the University of Copenhagen, where parts of this work were carried out. This research has been partially supported by Spain's Ministerio de Ciencia y Tecnolog\'{\i}a under programme AYA2\,002-082 (which includes FEDER funds). The work of the SAO team was supported by the Russian Astronomy Foundation and RFBR grant N01-02-17\,106a. KAP acknowledges partial support by RFBR grants 00-02-17\,164 and 99-02-16\,205. UL acknowleges partial support from Spain's Ministerio de Ciencia y Tecnolog\'{\i}a, under grant AYA 2\,002-03\,338, and Junta de Andaluc\'{\i}a. JPUF gratefully acknowledges support from the Carlsberg Foundation. This work was supported by the Danish Natural Science Research Council. This research has made use of NASA's Astrophysics Data System. The authors wish to thank Dr. G.B.Taylor for helpful comments during the refereeing process.

\end{document}